# THE COMBINATION OF CONVOLUTION NEURAL NETWORKS AND DEEP NEURAL NETWORKS FOR FAKE NEWS DETECTION

ZAINAB A. JAWAD *, AHMED J. OBAID

Department of computer science, University of Kufa College of Mathematical and Computer Science, Najaf 54001, Iraq.
*Corresponding Author: Zainaba.alhasani@student.uokufa.edu.iq,
ahmedj.aljanaby@uokufa.edu.iq.

**Abstract**

Nowadays, People prefer to follow the latest news on social media, as it is cheap, easily accessible, and quickly disseminated. However, it can spread fake or unreliable, low-quality news that intentionally contains false information. The spread of fake news can have a negative effect on people and society. Given the seriousness of such a problem, researchers did their best to identify patterns and characteristics that fake news may exhibit to design a system that can detect fake news before publishing. In this paper, we have described the Fake News Challenge stage #1 (FNC-1) dataset and given an overview of the competitive attempts to build a fake news detection system using the FNC-1 dataset. The proposed model was evaluated with the FNC-1 dataset. A competitive dataset is considered an open problem and a challenge worldwide. This system's procedure implies processing the text in the headline and body text columns with different natural language processing techniques. After that, the extracted features are reduced using the elbow truncated method, finding the similarity between each pair using the soft cosine similarity method. The new feature is entered into CNN and DNN deep learning approaches. The proposed system detects all the categories with high accuracy except the disagree category. As a result, the system achieves up to **84.6** % accuracy, classifying it as the second ranking based on other competitive studies regarding this dataset.

## 1. Introduction

The detection of fake news has recently stimulated the attention of the general audience and scientists as the spread of misleading information online grows, especially on the internet, such as newsfeeds, news articles, and newspaper articles. For example, according to a recent Jump-shot Tech Blog, on Facebook, 50% of the daily traffic represents fake news sites, while the remaining 20% present normal websites. Identifying fake content in online sources is crucial since 62% of U.S. adults read news on social media [1]. Until now, computational methods for detecting fake news have depended heavily on humorous news sources like "The Onion" and fact-checking websites like "Factcheck.org" and "Snopes." Moreover, using these publications presents several difficulties and possible consequences. For example, using humorous content as a source for misinformation can introduce underlying contributing effects into the analysis, such as jokes and illogicality. This is certainly relevant for humorous reports from "The Onion," which has previously been used to investigate other sentence characteristics such as jokes [2] and irony [3].

However, fact-checking websites are typically limited to a single area of interest, like politics. They necessitate human experience, making it challenging to find datasets that offer some level of generalization across several disciplines [4].

Consequently, recent work has explored how machine learning algorithms can extract linguistic features from textual articles. It can determine whether a fake news detection system is effective by evaluating how accurately it detects fake news. A set of datasets has been collected for this purpose, which can use to evaluate fake news detection systems, verify that the headline and body text match, identify mismatched sentences in the body text, and measure the social media spread of fake news [5].

In this work, we explain the main reason for choosing the FNC-1 dataset for evaluating the fake news detection system. Furthermore, this paper explains the main pre-processing methods used. On the other hand, it illustrates how the Elbow Truncated method was used for selecting the most effective feature. In addition, it briefly describes the feature generation step using TF-IDF. Moreover, it presents the cosine similarity method and demonstrates how it measures the similarity between documents. Finally, we illustrate the learning algorithm using CNN and DNN.

## 2.   Relate work

In this section includes a review of initial attempts to use the FNC-1dataset to build a misinformation news detection system. We focus on methods that combine Deep Learning (DL) techniques, Machine Learning (ML), and Natural Language Processing (NLP) techniques.

By suggesting the UCL Machine Reading (UCLMR) method, Riedel et al. 2017 [6] conducted an exciting experiment. The suggested method involves combining lexical and similarity information with DL approaches. This system offers simplicity in contrast to the most advanced Neural Network algorithms. Despite the simplicity of the implementation, the authors acquired an 81.72% FNC-1 score. The researchers believed their work would serve as a foundation for competing attempts on the FNC-1 dataset, although the already attained accuracy still needed improvement.

Another team, Gaurav Bhatt and his team 2017 [7], demonstrated a model created using DL methods and statistical, neural, and feature engineering heuristics. The suggested model was straightforward. Furthermore, it fulfils 83.08% of the accuracy, with higher accuracy in all instances expected disagree instances.

Benjamin Schiller et al. 2018 used a novel method for stance identification in this research [8] by applying memory networks with Convolutional Neural Networks (CNN), and Long Short-Term Memory (LSTM) approaches. A stance filtering component and the similarity-based matrix were used to enhance the model further. The suggested extension greatly enhances performance benefits and brings competitiveness. Results showed that the strategy effectively recognized prior knowledge; it scored 81.23%. The suggested model only manages one instance at a time and cannot deal with a collection of instances.

Borges et al. [9] presented a novel approach that combines Deep Neural Network (DNN) approaches with string external similarity features. The suggested model significantly improves the results of earlier research, which scored 82.23%. However, the authors reported that advanced sentence modelling techniques are required. Furthermore, instead of feed-forward approaches, the Recurrent Neural Network (RNN) method must be used for text encoding.

In another attempt by Robiert and his co-workers 2021[10], the researcher used RoBERTa as a foundation to improve the classification. In this case, two dense layers are included to reduce dimensions and, finally, the SoftMax classification layer.in addition, they were using soft cosine similarity. However, using summaries provided good results in this preliminary research, but the instances of agree and disagree still needed more improvement.

## 3. The proposed system

The main components of our proposed system are presented in this section. Figure 1 illustrates these components which described in detail.

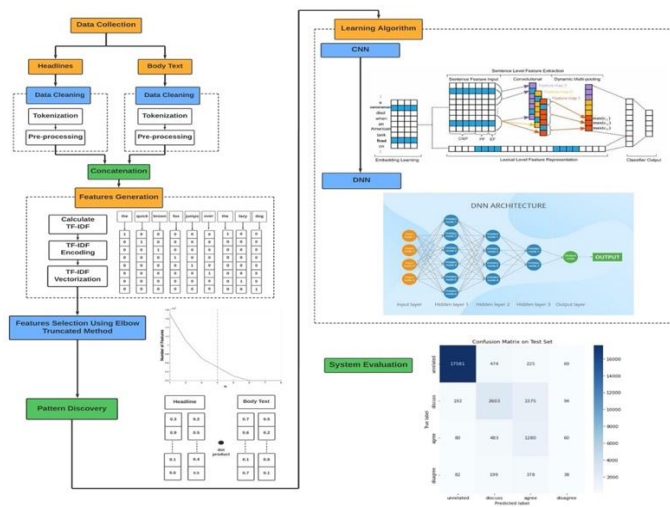

**Fig. 1: The Main Components of The Proposed System.**

We used the FNC-1 dataset in this search. The FNC-1 is a collective effort that includes over 100 volunteers and 71 teams from academia and industry worldwide. This dataset was designed for stance detection with 75,385 labeled headlines and articles. There are four label categories: agree, disagree, discuss, and unrelated. In the dataset, each headline is phrased as a statement. The goal of the FNC-1 is to explore how AI technologies could be employed to tackle fake news problems. They organize an annual competition to encourage the development of new tools that could assist human fact-checkers using ML, AI, and NLP.

After that, we used data cleaning. Raw data must be transformed into a format that can be understood and used. A real-world dataset is usually incomplete and has inconsistent formatting. It is possible to resolve such issues by preprocessing data and making it more efficient and complete for analysis. A successful data mining or machine learning project depends heavily on this process. As a result, knowledge can be discovered more quickly from datasets, which eventually affects the performance of machine learning models—the main steps of preprocessing, as shown in Fig. 2.

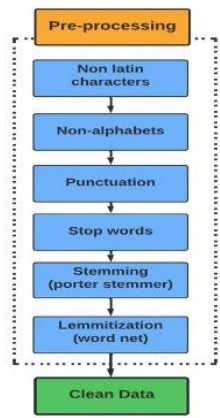

**Fig. 2: The Main Components of The Preprocessing.**

Among NLP tasks, text generation involves generating text with constraints like starting characters or starting words. In addition, generating text appears indistinguishable from human-written text. In the literature, this task is referred to as "natural language generation."

Text Generation is a branch of NLP. It uses mathematical linguistics and AI knowledge to automatically generate natural language texts that can fulfil certain communicative requirements. Which include:

**Calculate and Encoding by the (Term Frequency-Inverse Document Frequency) TF-IDF:** It is a statistical measure that determines how relevant a text is to a document in a list of documents. It is accomplished by multiplying two metrics: the number of times a term makes it appears in a document and the phrase's inverse document frequency across a list of documents. In this step, we perform the TF-IDF for the headline and the body text, which produce two encoding lists: one for the headline and the other for the body text. as shown in Fig. 3.

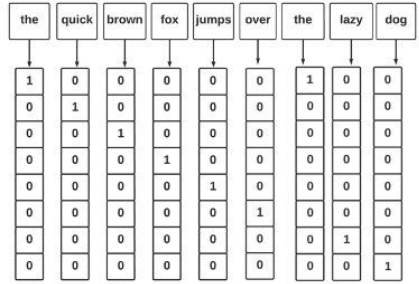

**Fig. 3: TF-IDF Vectorization Process**

Feature selection is crucial when building a fake news detection system for removing noisy features and keeping only those relevant to the system. After generating the TF-IDF vectors for both headlines in train stances and body text in train bodies, we use the Elbow truncated method to find the optimal number of features. Furthermore, we have 17000 concepts, while the optimal number of features after using this method is 5000. This step illustrates in Fig. 4.

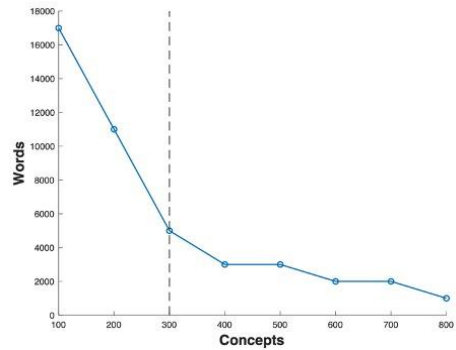

**Fig. 4: Feature Selection Process.**

After selecting the important feature, we contacted the headline in train stances and text bodies in train bodies in one CSV. This step is shown in Fig. 5.

|   | Headline | Body_Text |
|---|---|---|
| 0 | soldier shot parliament locked gunfire erupts ... | A small meteorite crashed into a wooded area i... |
| 1 | tourist dubbed 'spider man' spider burrows ski... | A small meteorite crashed into a wooded area i... |
| 2 | luke somers killed failed rescue attempt yemen | A small meteorite crashed into a wooded area i... |
| 3 | breaking soldier shot war memorial ottawa | A small meteorite crashed into a wooded area i... |
| 4 | giant 8ft 9in catfish weighing 19 stone caught... | A small meteorite crashed into a wooded area i... |

**Fig. 5: An example of merge process.**

Furthermore, it seeks to extract similar patterns from temporal data, such as periodic patterns. Discovering similar patterns is considered among the most crucial data-mining processes and may be used in various fields. Soft Cosine Similarity (SCM) allows us to assess the similarity between two documents in a meaningful way. The sentences have no words in common, but SCM can measure their similarity accurately by modelling synonymy. As part of the method, the documents are also represented as TF-IDF vectors (in other words, their frequency in the documents). It outperforms many state-of-the-art methods when applied to a semantic text similarity task within a community question-answering context. Thus, we use SCM in this study.

Finally, learning algorithms define the last level in building a fake news system. This procedure is essential for building a fake news system that can identify misleading news in real time. Different strategies, including ML classification, and DL approaches, are available for this purpose. As our main technique in this study, we used multiple deep learning classification algorithms, which include:

- Convolutional Neural Networks: This is a part of artificial neural networks commonly used in text mining. Compared to other text classification algorithms, CNNs require relatively little pre-processing. Unlike traditional algorithms, the network learns to optimize the filters (or kernels) through automated learning, whereas in traditional algorithms, these filters are designed by hand. This approach is advantageous because it is independent of prior knowledge and human intervention when obtaining features.

- Deep Neural Networks: In DNNs, data flows from the input layer to the output layer without looping back. A DNN begins by creating a map of virtual neurons and assigning random numerical values to connections between them. A combination of weights and inputs results in output between 0 and 1. An algorithm would adjust the weights if the network failed to recognize a particular pattern accurately. As a result, the algorithm can gradually increase the importance of certain parameters until it finds the correct mathematical manipulation to process the data fully.

After finding the SCM between headlines and body text, we add the stance column with four labels (agree, disagree, discuss, and unrelated), which will represent our dependent variable. In addition, we enter these columns into CNN learning models with some properties, as shown in Table 1.

**Table 1: Properties of CNN learning model.**

| CNN layer | Dense | Activation |
|-----------|-------|------------|
| Input layer | 1024 | relu |
| Hidden layer | 128 | relu |
| Output layer | 4 | SoftMax |

On the training CSV, with the batch size being 512 and the number of epochs being 80, the best result with a categorial accuracy (average accuracy of stances) of 0.997 and loss function of 0.659 because the first ten epochs were over-fitting. We compute the confusion matrix as shown in Fig. 6.

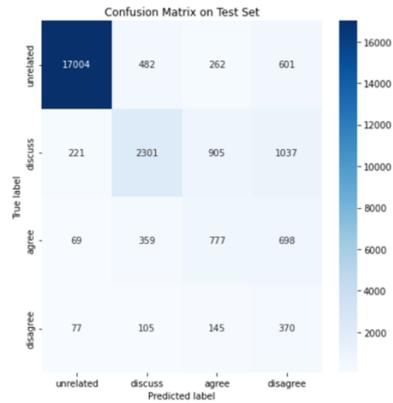

**Fig. 6: The Confusion Matrix of CNN Algorithm.**

That is challenging on the competition's CSV, which has real-time news and does not exist in training or testing CSV.

After we learn the training CSV, we do the same Learning on the competition CSV. The best result was validation-categorial accuracy (average accuracy of stances) of 0.805 and validation-loss function of 0.938 because the first ten epochs were over-fitting. Furthermore, we compute the confusion matrix as shown in Fig.7. Finally, we loaded the weighted CNN to the DNN model to enhance our results, which increased the accuracy to 84.6%.

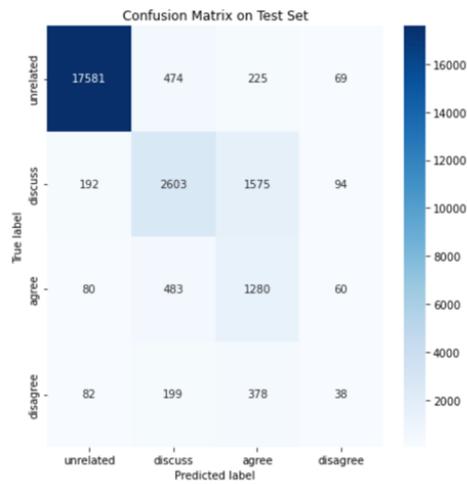

**Fig. 7: The Confusion Matrix of DNN Algorithm.**

We used this metric for evaluating classification models that specifically measures the similarity between predicted and actual fake news. Furthermore, it is possible to calculate using an equation. The number of documents detected by the system is denoted by True Positive (TP), while the number of correctly predicted fake documents is represented by True Negative (TN). Furthermore, False Positive (FP) represents the number of real documents classified as fake by the system. Finally, False Negative (FN) refers to the number of fake documents classified as real by the system.

The F-score is an important metric that reflects the success of detecting fake news rather than real news. Precision represents the percentage of correctly classified observations as a fake document. Sensitivity, in contrast, is defined as the ratio of real news classified to actual real news [11]:

- **Precision** $= \frac{|TP|}{|TP| + |FP|}$ **(1)**

- **Recall** $= \frac{|TP|}{|TP| + |FN|}$ **(2)**

- **F1-Score** $= \frac{2 * precision * Recall}{precision + Recall}$ **(3)**

- **Accuracy** $= \frac{|TP| + |TN|}{|TP| + |TN| + |FN| + |FP|}$ **(4)**

These metrics are widely used in the ML and DL community and allow to evaluate a classifier's performance from various angles.

## 4. Results and Discussion

System implementation was performed on a Macintosh platform with a dualcore Intel Core i5 processor and 16 GB of memory, illustrated in Table 2.

Both Microsoft Visual Studio 2019 and MATLAB R2020b were used. As shown in Table 3, pre-processing takes 12 seconds on average, converting the data into a TF-IDF vectorizer takes 16 minutes, and calculating SCM between headlines and body text takes 24 minutes. The test and the training of CNN take 5.55 minutes, respectively. On DNN, the system can be applied in real-time in 2.034 seconds, demonstrating its ability to be applied in real-time.

The proposed system is evaluated via two classifiers based on the features obtained using Elbow truncated algorithms. CNN and DNN are these classifiers. Based on the CNN classifier, which is most commonly used to analyse or understand natural language processing, the two classifiers perform well in detecting fake news.

**Table 2: Summary of the Experiment Results.**

| Classifier | Accuracy | Precision | Recall |
|---|---|---|---|
| Proposed Model (CNN, DNN) | 84.6% | 84.6% | 84.6% |

**Table 3: Summary of the execution time.**

| Model | Execution time |
|---|---|
| Pre-processing | 12 sec |
| TF-IDF | 16 min |
| SCM | 24 min |
| CNN | 5.55 min |
| DNN | 2 sec |
| Totally | 45.69 min |

Furthermore, we calculate the results of each stance in the dataset, as shown in Table 4.

**Table 4: Classification Result of Proposed System based on FNC-1 Stances**

| Stance | Accuracy | Precision | Recall | F-score |
|--------|----------|-----------|--------|---------|
| Unrelated | 95.04 | 98.03 | 95.81 | 96.91 |
| Discuss | 87.70 | 69.25 | 58.31 | 63.31 |
| Agree | 88.47 | 37.02 | 67.26 | 47.75 |
| Disagree | 96.00 | 05.45 | 14.56 | 07.93 |

Across all stances except disagree, we achieved higher scores and class-wise accuracy, as shown in Fig. 8. It is a problem since the importance of disagree is similar to the importance of agree and discuss. Not only are there a relatively small number of news pairs in the disagree category, but they also contain divergent news articles. Deep models, including top teams, do not perform well when classifying disagree labels.

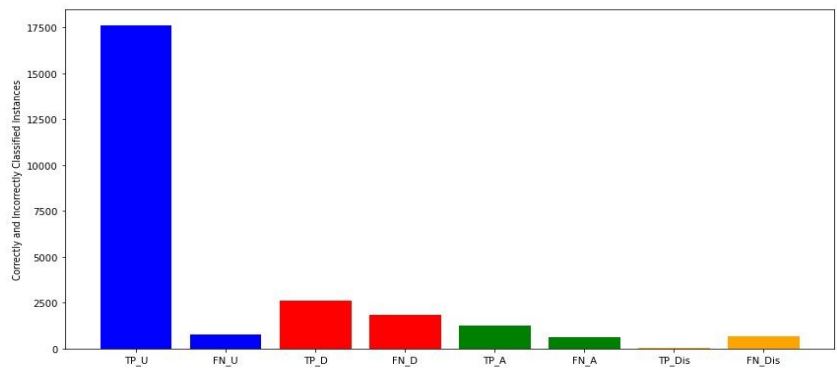

**Fig. 8: The Results of Propose System Based on Evaluation Metrics**

Moreover, we compared the proposed system's performance to previous works based on metrics of accuracy and methods used. Table (5) shows that our proposed approach outperforms most previous approaches. For example, [6] used the ULCMR method to achieve 81.72% accuracy, while [7] used statistical, neural, and feature engineering heuristics to achieve 83.08% accuracy. In another example, the authors in [8] used Novel memory network enhancements with CNN and LSTM to achieve 81.23% accuracy. Although in [10], the researcher achieved 90% accuracy. On the other hand, the proposed system achieved 84.6% accuracy and needs to combine CNN and DNN with SCM.CNN and DNN are used to demonstrate the ability of the system to detect fake news, as shown in Fig. 9.

**Table 5: Comparison Between the Proposed Approach with Previous Works Based on State of the Art**

| Reference Number | Year of Study | Technique | Accuracy |
|------------------|---------------|-----------|----------|
| [6] | 2017 | UCL Machine Reading's (UCLMR) system. | 0.8172 |

| [7] | 2017 | Statistical, neural and feature engineering heuristics. | 0.8308 |
|---|---|---|---|
| [8] | 2018 | Novel memory network enhances with CNN and LSTM. | 0.8123 |
| [12] | 2018 | Rich stacked LSTM approach. | 0.8020 |
| [9] | 2018 | Combining deep neural network approaches with string similarity features. | 0.8223 |
| [13] | 2020 | Based on the core concept of Credibility Reviews (CRs). | 0.7160 |
| [10] | 2021 | DL (SoftMax activation) with SCM and RoBERTa. | 0.9073 |
| Propose system | 2022 | Combine CNN and DNN with SCM. | 0.8460 |

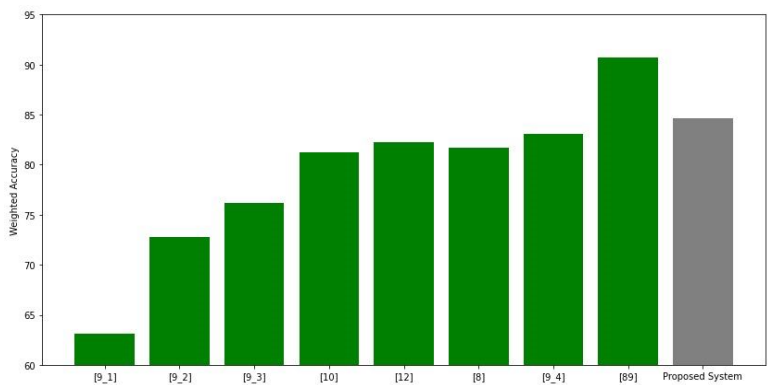

**Fig. 9: Comparison between the proposed approach with previous works Based on State of the Art**

Figure 10 illustrates the purpose system with the other previous works in state of the art, which is higher than other proposed models.

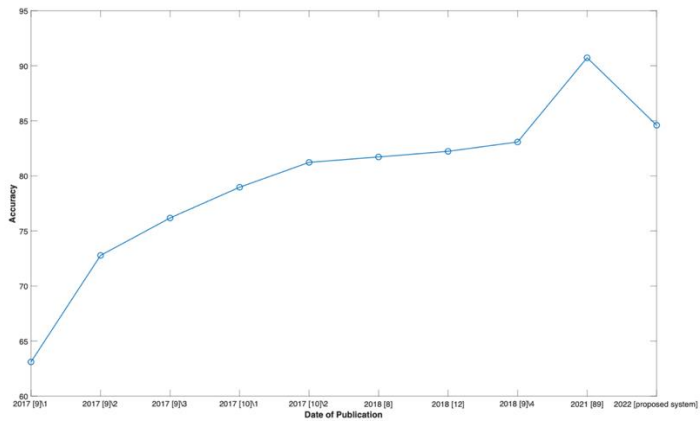

**Fig. 10: Comparison between the proposed approach with state the art.**

# 5. Conclusions

Due to social media's increasing popularity, more and more people are consuming news via social media instead of traditional media. In addition, social media has been used to spread fake news, which can negatively impact individual users and society. As part of this research, we reviewed existing literature on the characterization and detection of fake news. Researchers attempted different approaches to build fake news detection systems using machine learning and deep learning approaches using natural language processing. This work focused on the competitive studies that used the FNC-1 dataset. From studying the existing works, it is concluded that the proposed approaches fail to build a system that can detect fake news efficiently.

The current work proposed a detection system to detect fake news in the FNC-1 competitive dataset. The proposed system implies cleaning the data, extracting new features using different natural language processing techniques, and reducing the number of features using the elbow truncated method. Furthermore, we found the similarity between each pair of headline and body text, and the resulting feature is entered into CNN and DNN learning approaches. As a result, the system can recognize if the headline agrees with the body text, discuss it, disagree with it, or they are unrelated.

The systems give up to 84.6% accuracy, where it detects all these cases with high accuracy except the disagree case, and the main reason for that is that the training pairs of this category are not frequent. As a result, the proposed system can be considered the second-ranking among the competitive studies with the FNC-1 dataset, where it outperforms most of the previous works in terms of accuracy.

# References

[1] Gottfried, J. and E. Shearer, *News use across social media platforms 2016.* 2019.

[2] Mihalcea, R. and C. Strapparava. *Making computers laugh: Investigations in automatic humor recognition.* in *Proceedings of Human Language Technology Conference and Conference on Empirical Methods in Natural Language Processing.* 2005.

[3] Wallace, B.C., *Computational irony: A survey and new perspectives.* Artificial intelligence review, 2015. **43**(4): p. 467-483.

[4] Pérez-Rosas, V., et al., *Automatic detection of fake news.* arXiv preprint arXiv:1708.07104, 2017.

[5] Zhou, X. and R. Zafarani, *A survey of fake news: Fundamental theories, detection methods, and opportunities.* ACM Computing Surveys (CSUR), 2020. **53**(5): p. 1-40.

[6] Riedel, B., et al., *A simple but tough-to-beat baseline for the Fake News Challenge stance detection task.* arXiv preprint arXiv:1707.03264, 2017.

[7] Bhatt, G., et al., *On the benefit of combining neural, statistical and external features for fake news identification.* arXiv preprint arXiv:1712.03935, 2017.

[8] Mohtarami, M., et al., *Automatic stance detection using end-to-end memory networks.* arXiv preprint arXiv:1804.07581, 2018.

[9] Borges, L., B. Martins, and P. Calado, *Combining similarity features and deep representation learning for stance detection in the context of checking fake news.* Journal of Data and Information Quality (JDIQ), 2019. **11**(3): p. 1-26.

[10] Sepúlveda-Torres, R., et al. *Exploring summarization to enhance headline stance detection*. in *International Conference on Applications of Natural Language to Information Systems*. 2021. Springer.

[11] Shu, K., et al., *Fake news detection on social media: A data mining perspective.* ACM SIGKDD explorations newsletter, 2017. **19**(1): p. 22-36.

[12] Hanselowski, A., et al., *A retrospective analysis of the fake news challenge stance detection task.* arXiv preprint arXiv:1806.05180, 2018.

[13] Denaux, R. and J.M. Gomez-Perez. *Linked credibility reviews for explainable misinformation detection*. in *International Semantic Web Conference*. 2020. Springer.